\documentclass[]{spie}  

\newcommand{\code}[1]{\texttt{#1}}

\usepackage{amsmath,amsfonts,amssymb}
\usepackage{graphicx}
\usepackage[colorlinks=true, allcolors=blue]{hyperref}
\usepackage{multirow}
\usepackage{graphicx}
\usepackage{lscape}
\usepackage{caption}
\usepackage[table,xcdraw]{xcolor}
\usepackage{orcidlink}
\usepackage{amsmath}

\title{The Simons Observatory: Alarms and Detector Quality Monitoring}

\author[1]{David V. Nguyen\,\orcidlink{0000-0002-7575-8145}}
\author[1]{Sanah Bhimani\,\orcidlink{0000-0002-9763-1663}}
\author[2,3]{Nicholas Galitzki\,\orcidlink{0000-0001-7225-6679}}
\author[1]{Brian J. Koopman\,\orcidlink{0000-0003-0744-2808}}
\author[1]{Jack Lashner\,\orcidlink{0000-0002-6522-6284}}
\author[1]{Laura Newburgh\,\orcidlink{0000-0002-7333-5552}}
\author[1]{Max Silva-Feaver\,\orcidlink{0000-0001-7480-4341}}
\author[4,5]{Kyohei Yamada\,\orcidlink{0000-0003-0221-2130}}

\affil[1]{Wright Laboratory, Department of Physics, Yale University, New Haven, CT 06511, USA}
\affil[2]{Department of Physics, University of Texas at Austin, Austin, TX, 78712, USA}
\affil[3]{Weinberg Institute for Theoretical Physics, Texas Center for Cosmology and Astroparticle Physics, Austin, TX 78712, USA}
\affil[4]{Joseph Henry Laboratories of Physics, Jadwin Hall, Princeton University, Princeton, NJ 08544, USA}
\affil[5]{Department of Physics, The University of Tokyo, Tokyo 113-0033, Japan}

\authorinfo{Corresponding author: David V. Nguyen\\ E-mail: david.nguyen@yale.edu}

\begin{document} 
\maketitle

\begin{abstract}
The Simons Observatory (SO) is a group of modern telescopes dedicated to observing the polarized cosmic microwave background (CMB), transients, and more. The Observatory consists of four telescopes and instruments, with over 60,000 superconducting detectors in total, located at \(\sim \)5,200\,m altitude in the Atacama Desert of Chile. During observations, it is important to ensure the detectors, telescope platforms, calibration and receiver hardware, and site hardware are within operational bounds. To facilitate rapid response when problems arise with any part of the system, it is essential that alerts are generated and distributed to appropriate personnel if components exceed these bounds. Similarly, alerts are generated if the quality of the data has become degraded. In this paper, we describe the SO alarm system we developed within the larger Observatory Control System (OCS) framework, including the data sources, alert architecture, and implementation. We also present results from deploying the alarm system during the commissioning of the SO telescopes and receivers.

\end{abstract}


\renewcommand{\thefootnote}{\fnsymbol{footnote}}

\section{Introduction}
Measurements from CMB experiments have resulted in precise constraints on $\Lambda$CDM parameters and derived quantities. \cite{planck} The Simons Observatory (SO) is a ground-based CMB experiment being commissioned in the Atacama Desert in Chile. It consists of three 0.5\,m diameter small-aperture telescopes (SATs) and one 6\,m diameter large-aperture telescope (LAT), totaling over 60,000 antenna-coupled, multi-chroic transition edge sensor (TES) bolometers across a frequency range of 27-280\,GHz. \cite{galitzki2024simons, 
Zhu_2021, Gudmundsson_2021, McCarrick_2021} The detectors use a microwave multiplexing ($\mu$MUX) readout architecture and SLAC Microresonator Radio Frequency (SMuRF) readout electronics. \cite{smurf, Kernasovskiy_2018} The increased detector count provides improved sensitivity compared to previous generations of telescopes. As a result, SO will improve our understanding of science goals including theories of inflation, the number of light relic particles, lensing of large-scale structure, the kinetic and thermal Sunyaev-Zel'dovich effect in galaxy clusters, extra-galactic point sources, and transient events. \cite{SO_Goals}


To achieve this increased detector count across a wide variety of science cases, SO has deployed detectors across four separate telescopes. The resulting telescope platforms, millimeter receivers, and site infrastructure include over 5,000 slow data fields across cryogenics, power distribution, computing, networking, weather, etc. We call the non-detector data housekeeping (HK) data, acquired slower than the detector sampling rate. To perform control, data acquisition, and monitoring of HK systems across the SO, we have developed the Observatory Control System\footnote{\url{https://github.com/simonsobs/ocs}} (\code{ocs}). \cite{ocs} A key component of observations is the monitoring and alarming based on metrics from these subsystems. The alarm system must be able to monitor the health of the entire telescope, provide quick overviews of the state of the system, emit alerts that sufficiently describe the issue, and notify researchers via various methods depending on priority level and user preference. It must also be easily modified and scalable as new alarms continue to be added with more subsystems and additional telescopes. Using the data feeds monitored by \code{ocs}, we use \code{campana} to generate alarms based on Grafana alert rules and to distribute these alerts using various notification methods. These alarms are monitored daily to ensure proper observations. \par


In this paper, we present data sources for alarms in Section~\ref{data sources for alarms}, including both HK and detector data. In Section~\ref{alarm system overview}, we describe an overview of the alarm system, detailing its requirements, architecture, and implementation. Next, in Section~\ref{deployment}, we describe how the alarm system is deployed at the site in Chile, improvements and successes, and plans moving forward. Finally, we conclude in Section~\ref{summary}. Appendix~\ref{appendix:a} contains the table of acronyms used within this paper. \par


\section{Data Sources for Alarms} \label{data sources for alarms}

SO has a large collection of data sets for assessing the health of the telescopes, site equipment, and other subsystems, along with tracking the detector data quality for science analysis. From these housekeeping and detector-related data sources, we create alarms depending on user-defined conditions (e.g., safe ranges, binary states, or combinations of thresholds). As of this writing, SO acquires data from over 5,000 fields; we use \textgreater500 of those fields to generate alarms at various levels. We will describe these data sources in this section.\par 


\subsection{Housekeeping (HK) Data} \label{HK data}


HK data sets come from any hardware devices except those for detector data acquisition. In SO, we typically divide into 2 broad categories: telescope-specific and the site. To target the most relevant researchers when an HK system is in an alarm state, we separate the alarms into the following groups (an example of an alarm state for most of these groups is given in Table~\ref{table:alerts}):

\begin{itemize}
    \item Computing. These HK data fields come from computers at the site and are usually acquiring data such as disk usage, CPU usage, and memory usage (via Telegraf\footnote{\url{https://www.influxdata.com/time-series-platform/telegraf/}} instead of \code{ocs}).
    \item SMuRF. \cite{smurf, Kernasovskiy_2018} These HK data fields track the detector readout system health. These include board temperatures, board current draw, and coolant leak sensors.
    \item Cryogenics. These HK data fields acquire data from the dilution refrigerator (DR), which cools down the detectors to superconducting temperatures, and associated cryostat sources. These include DR temperatures, pressures, and flow; compressor state, temperatures, and pressure; cryostat pressure and temperatures.
    \item Half-Wave Plate. Each SAT has a cryogenic half-wave plate that spins at $\sim$2\,Hz to modulate the incident polarization. \cite{hwp} HK data fields from this subsystem include rotation angle, rotation speed, and IRIG (absolute) timing.
    \item Power. These HK data fields acquire data from power storage systems and generators. These include uninterruptible power supply (UPS) health (e.g., battery state, charge remaining) and diesel generator health (e.g., fuel level, shutdown status, electrical trip status).
    \item Platform. These HK data fields acquire data from the telescope movable platform and its antenna control unit (ACU) which drives the platform. These fields include position and velocity, mode state (safe or remote), and time synchronization.
    \item Agents. As described in Section~\ref{architecture and dependencies}, HK agents communicate between hardware devices and the overall OCS software architecture. Monitoring for the agent operation status catches agent crashes.
    \item Timing. The SO timing system is centralized; timing is distributed either over the network as PTP or over dedicated fiber as specially generated timing signals defined by the SMuRF systems for the detector timing. \cite{brian_spie} The central timing device and the edge-clocks that receive timing signals for calibrators on the platforms have HK data fields such as GPS synchronization state, PTP state, and PTP accuracy.
    \item Environmental. These HK data fields are used to monitor the state of the site conditions. These include weather metrics such as wind speed, temperature, and precipitable water vapor (PWV).
    \item Remote Observing Schedule. SO has a scheduler system that runs a set of commands in order of line number. \cite{yilun_spie} We monitor whether that system has resulted in a fault and is no longer observing.
\end{itemize}

Some metrics are more important to monitor than others, as abnormal conditions may cause hardware damage or safety concerns. The alarm system distributes notifications appropriately depending on the severity level of alarms. At the time of writing, four HK metrics trigger phone calls (in addition to other notification methods), two of which are in Table~\ref{table:alerts}. These are: 
\begin{itemize}
\item DR temperature: During nominal observations, if the 100\,mK stage exceeds 120\,mK, the cryogenic system is in a critical state and needs immediate attention. 
\item Pulse Tube Cryocoolers (PTCs) state: PTCs are used to cool the DR and also as cryostat radiation shields for all receivers. PTC shutdowns require immediate attention.
\item Wind speed: Wind speed is important because it determines whether it is safe for people to work at the site or for the telescope platforms to move. A phone call is generated if gusts are \textgreater~70\,km/hour since the platforms cannot observe during those conditions. Note: a separate alert is distributed via other notification methods at \textgreater~50\,km/hour for site personnel safety, who often do not have access to phone calls due to reception at the site.
\item Sun avoidance: A phone call is triggered if the boresight of the telescope is within some distance of the sun. The exact degree value depends on the telescope. Solar avoidance has been implemented within the platform control; however, under certain conditions (e.g., manual control of telescope) pointing too close to the sun is still possible.
\end{itemize}

\subsection{Detector Quality Metrics}

Detector data is important to monitor since it provides the most direct assessment of observation quality. There are two main detector data quality sources for the alarms. The first source uses SMuRF HK feeds: in particular, data feeds that track the number of detectors in the superconducting transition (i.e., in a usable state for CMB observations). If too few detectors are on transition, we trigger an alarm under the assumption that the automatic detector biasing routine failed in some way. \par

The second source of alarms for detector data quality results from the data processing pipeline. The processing pipeline reads in the raw detector data, performs data quality flags based on common response to the sky and the HWP signal (for the SATs), and finds and corrects glitches; this produces an output required for quick-look maps \footnote{\url{https://github.com/simonsobs/sotodlib}}. The processing is run at the site separately for each telescope within 12 hours of data acquisition and is automated using \code{prefect}\footnote{\url{https://www.prefect.io/}}. \cite{yilun_spie} The same alarm system described in Section~\ref{alarm system overview} is used to alert on both metrics produced from the data processing pipeline \textit{and} the HK fields described above. This is useful for catching otherwise unnoticed issues, not for real-time application. For example, we can be notified if too many detectors are removed during a script that cuts detectors due to some criteria. This may lead us to believe that detector calibration was unsuccessful. \par

\begin{figure}
    \centering
    \includegraphics[scale=0.5]{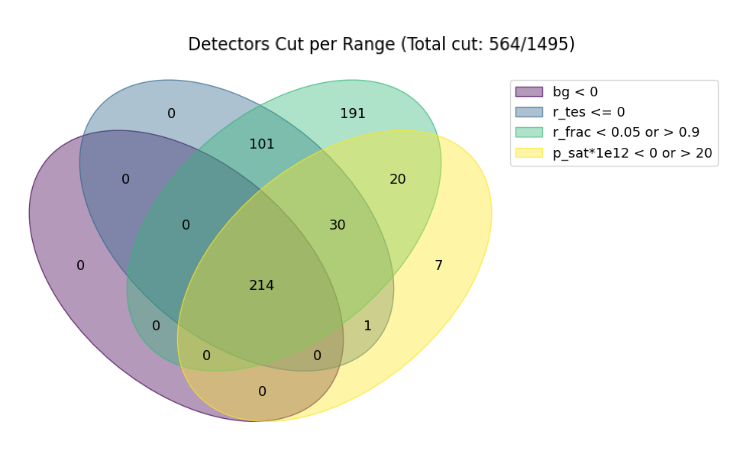}
    \caption{An example 4-set Venn diagram to demonstrate the detector bias cuts. Each set is a parameter that causes the detector to be removed. The bias group (bg) is flagged if \textless0 since unassigned detectors are labeled with -1. The detector resistance ($r_{tes}$) is flagged if \textless=0. The fractional resistance ($r_{frac}$) is flagged if outside of the defined range (0.05, 0.9). The saturation power ($p_{sat}$) is flagged if outside of the defined range (0, 20). Overlaps indicate detectors that are cut in common between different parameters. This is an example detector wafer from commissioning of one of the SATs, in which there are less desirable observing conditions to illustrate the cut effects. In this case, 564 out of 1495 total detectors were cut, 214 of which were common to all flags. This visualization indicates that the $r_{frac}$ cut is the most aggressive, and all detectors removed through the $r_{tes}$ and bg cuts are in common with other parameters, implying these are less discriminatory.}
    \label{fig:bias_cuts_venn}
\end{figure}

The pipeline also produces visualizations at each step, showing the transformed signal and related statistics, which aids inspection to ensure that data quality is adequate for CMB maps. For example, at the start of each observation, bias steps are taken to measure the response from small steps in detector bias voltage. Several parameters can be deduced from this procedure: the location of the detector in its superconducting transition ($r_{frac}$), the estimated TES resistance ($r_{tes}$), if detectors could not be assigned to a bias group due to bad fit, and the number of saturated (i.e., non-responsive) detectors. We remove detectors whose bias parameters are outside the expected range. These cuts can be implemented separately; however, they frequently overlap because they can probe the same poor behavior in the transition (e.g., saturated channels because of higher loading when the water vapor content in the atmosphere is large). To visualize this process to help understand how many detectors are cut from each parameter, we plot their overlap in Figure~\ref{fig:bias_cuts_venn} as a 4-set Venn diagram. Generating this figure is included in the automated processing scripts; this allows us to rapidly assess behavior in conjunction with alarms. \par



\section{Alarm System Overview} \label{alarm system overview}
The SO alarm system is a collection of software packages intended to monitor the observatory and alert observers of critical errors. Different groups of people such as Remote Observing Coordinators (ROC), site engineers, and system expert personnel can receive alerts. ROCs are researchers who take shifts to monitor the status of each telescope and respond to alarms. ROCs efficiently receive alerts, facilitating immediate response to recover systems for regular operations. This system was designed to allow ease of use through rapid inspection and multiple methods of receiving alerts. In this section, we describe the requirements needed for successful operations of the alarm system as well as the architecture and implementation. \par

\begin{figure}
    \centering
    \includegraphics[scale=0.4]{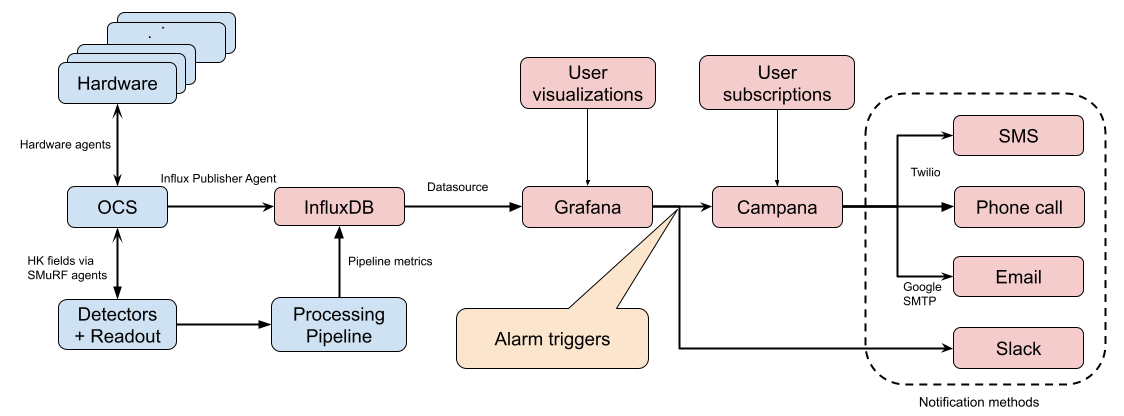}
    \caption{The SO alarm system architecture. \code{ocs} Agents monitor hardware and SMuRF HK metrics, sending data feeds to InfluxDB (via the Influx Publisher Agent). The data processing pipeline also produces detector data quality metrics, which are sent to InfluxDB. Using the InfluxDB as the data source, Grafana monitors the data feeds and emits an alert when thresholds are surpassed. Users create visualizations and define alert rules on Grafana. When an alarm triggers, alerts are sent to Slack as well as \code{campana}, which sends emails, SMS, and phone calls. Users subscribe to \code{campana} to receive alerts. Nodes colored blue are data sources as described in Section~\ref{data sources for alarms} and those colored red consist of the architecture and contact points described in Section~\ref{alarm system overview}.}
    \label{fig:alarm_architechture}
\end{figure}

\subsection{Requirements}
The alarm system must be able to monitor the health of every component, both hardware and software, across all telescopes and the site itself. The state of the system should be easily discernible by any SO member, whether that may be the ROC, site engineer, or system expert. When issues arise, the system must emit alerts that clearly describe the problem and contain a link to the live monitor for the data which sourced the alarm. Additionally, the system must emit alerts that link to failed observation schedules. These alerts should be distributed to the appropriate people who can best respond to the situation; thus, the system should accommodate various notification methods and emit alerts according to their defined group and severity level. It must also be scalable since new alarms will continue to be added with additional subsystems and telescopes. \par

\subsection{Architecture and Implementation} \label{architecture and dependencies}


The architecture and interfaces of the SO alarm system, as well as software dependencies, are shown in Figure~\ref{fig:alarm_architechture}. This system is described in more detail in this section. \par

\begin{figure}
    \centering
    \includegraphics[scale=0.4]{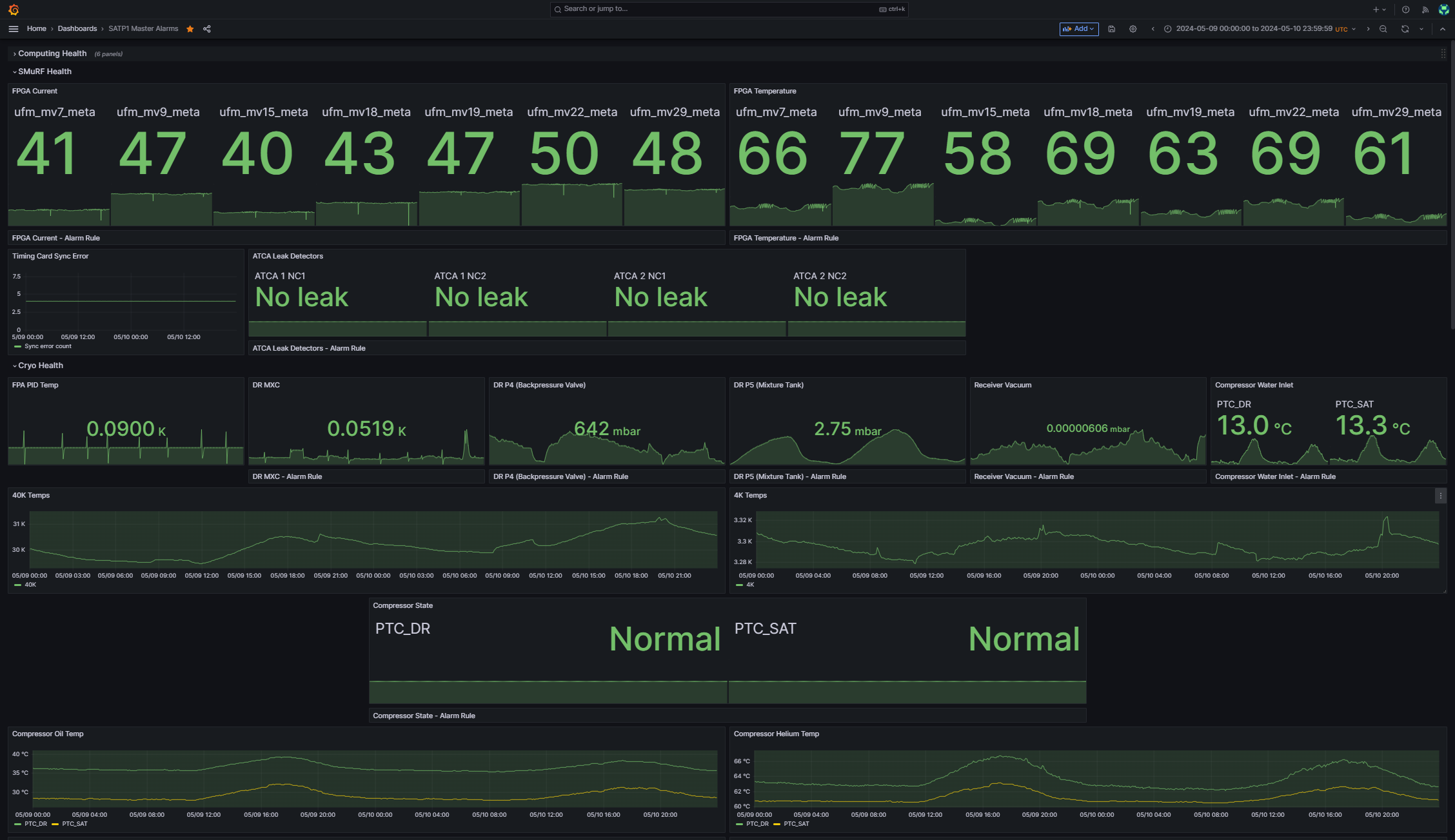}
    \caption{An example Grafana dashboard used to monitor metrics for one of the SATs. In this case, we show the health of the SMuRF system and the cryogenics system. This dashboard has a combination of different display types: time-series plots of temperatures, some with multiple sensors shown in the same plot, as well as `single-stat' panels, which give the most recent value (for this case, the FPGA currents/temperatures and DR temperatures/pressures). Also shown are the single-stat states for binary flags from systems (in this case, that no leak is present in the system). Here, panels show green for normal operations. When thresholds are exceeded and alarms are triggered, the panels show red. Panels are grouped by the alarm groups described in Section~\ref{data sources for alarms}.}
    \label{fig:alarm_dashboard}
\end{figure}

\begin{figure}
    \centering
    \includegraphics[scale=0.4]{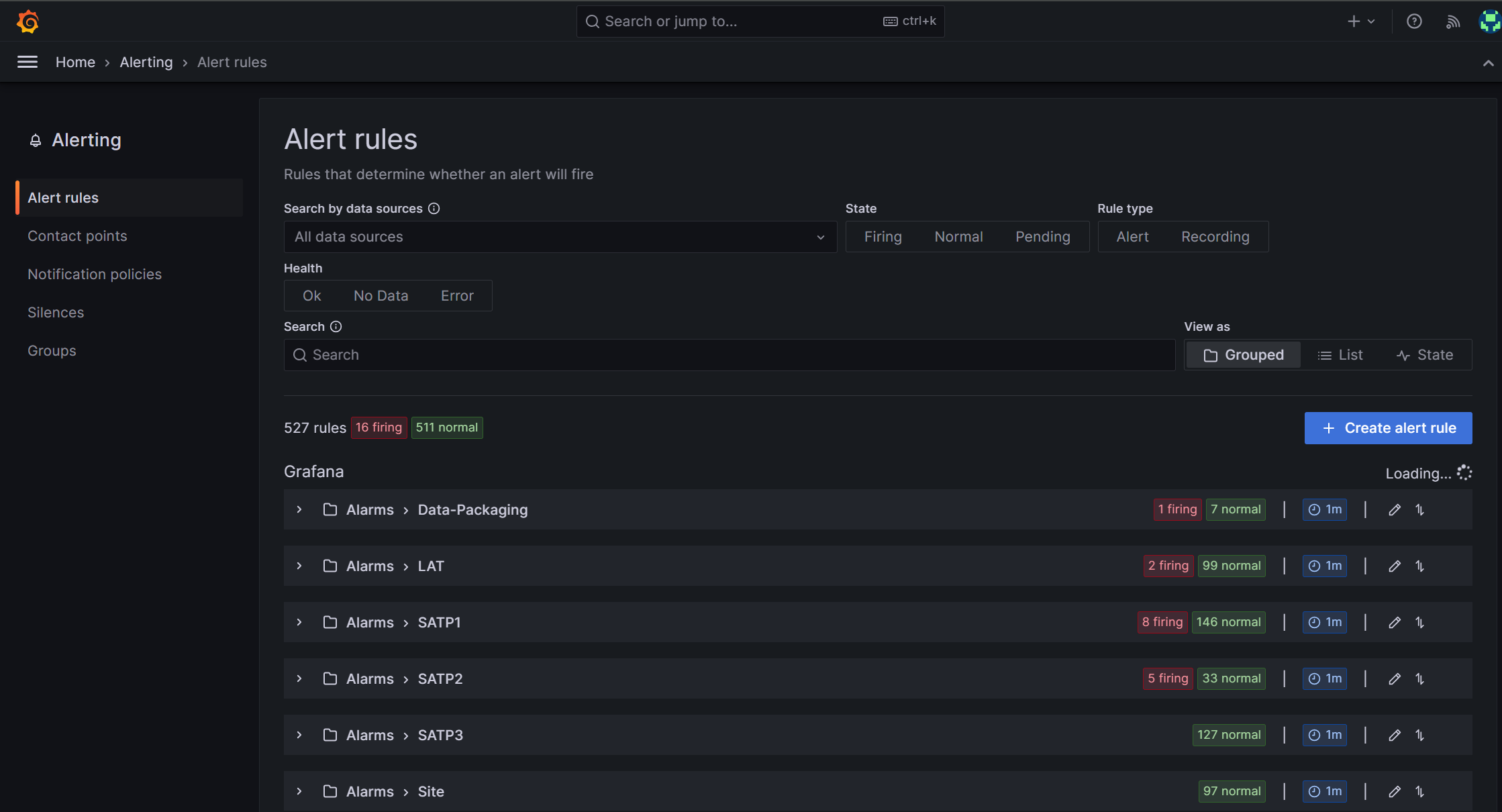}
    \caption{The Grafana alert rules page shows all active alerts grouped by the alarm overview dashboards. The list can be filtered to show currently firing alerts.}
    \label{fig:alert_rules}
\end{figure}

\subsubsection{Visualization and Alert Generation}
The first component of the alarm system is the interface between hardware and the \code{ocs}. The \code{ocs} is the central software stack that monitors and controls the telescope and site hardware via \code{ocs} Agents. \cite{ocs} Agents are individual software servers, each connecting to a different piece of hardware. Agents are written in Python and typically deployed using Docker\footnote{\url{https://www.docker.com/}} containers with mounted configuration files. As noted in Section~\ref{data sources for alarms}, some examples of metrics monitored by Agents include the temperature of the DR, the PWV levels from the radiometer, the fuel level of the diesel generators that power the entire observatory, and some SMuRF health metrics such as number of detectors that are on their superconducting transition after a bias step. \par

\code{ocs} sends HK data to InfluxDB\footnote{\url{https://www.influxdata.com/}}, a time series database, via the Influx Publisher Agent. The real-time detector data processing pipeline also produces metrics that are published to InfluxDB. Grafana\footnote{\url{https://grafana.com/}} uses the InfluxDB as the data source to visualize this time series data, whether from \code{ocs} Agents, Telegraf, or the processing pipeline. Grafana is a web application used to visualize and analyze time series data through the use of ``dashboards". A dashboard consists of ``panels'', each displaying the data in a user-specified way. The panels can be arranged to group subsystem components together (e.g., as described in Section~\ref{data sources for alarms}). An example of part of an SO dashboard is shown in Figure~\ref{fig:alarm_dashboard}, with more detail provided in the caption. We lay out these dashboards to optimize space and give the most crucial information at a glance. The example shown in Figure~\ref{fig:alarm_dashboard} is an alarm overview page for one of the SAT telescopes and contains a subset of the most essential data to capture the health of the SAT. Additional dashboards can be created by users for specific purposes (e.g., investigating hardware issues); however, only the alarm overview dashboards are used to define the alarms. There are five alarm overview dashboards: one for each of the 3 SATs, one for the LAT, and one for the site. The panels show green or red depending on defined thresholds to illuminate which subsystem is in an error state. These thresholds are defined in Grafana via ``alert rules''; all alert rules can be viewed on a single page on Grafana, providing an overview of the alarm states of all observatory systems (Figure~\ref{fig:alert_rules}). Since Grafana alert rules are easily duplicated, setting new alerts is easy and can be centralized to one person to avoid miscommunication. Grafana generates the alarms used by SO based on these alert rules. Grafana's ``contact points'' determine where alert notifications are sent. \par

\subsubsection{Alert Contact Points} \label{alert contact points}
SO uses two primary contact points: Slack and \code{campana}. The SO collaboration uses Slack as a primary messaging tool and an essential part of coordinating observatory operations; sending alerts via Slack allows for user-friendly notification. Each group of alarms is connected to a separate Slack channel using webhooks pushed from Grafana to Slack. Each user can join channels relevant to systems they want to monitor to receive notifications and respond accordingly. \par

\begin{figure}
    \centering
    \includegraphics[scale=0.4]{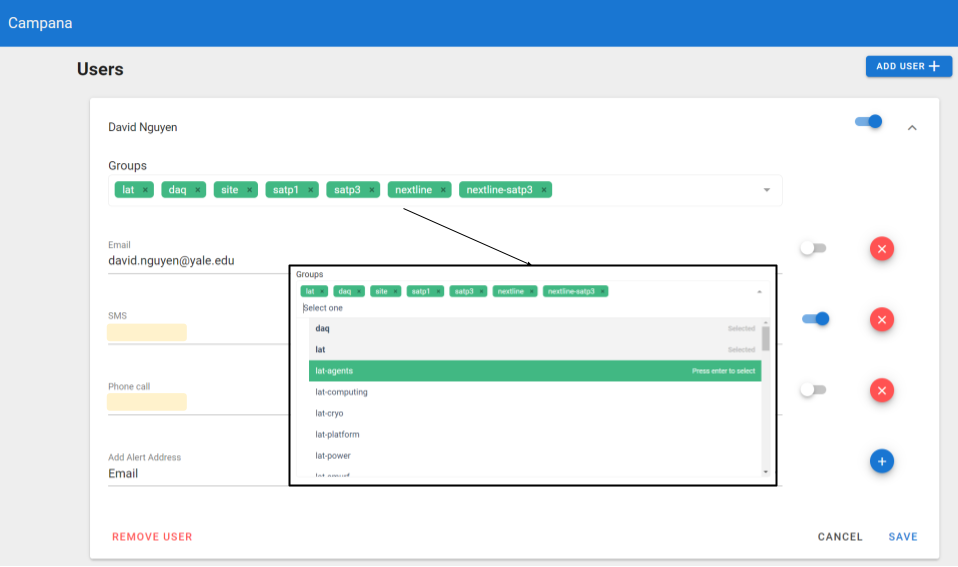}
    \caption{The \code{campana} webpage where ROCs can enter their information, subscribe to alert groups, and toggle active notification methods. The inset shows the drop-down menu of groups to select. In this example, the user would receive text messages, but not emails and phone calls.}
    \label{fig:campana}
\end{figure}

Slack receives all alerts, regardless of priority level. Since any of the \textgreater500 possible alerts can be sent to Slack, the Slack channels can become congested; thus, ROCs can easily miss notifications or develop notification fatigue. To mitigate these issues, we developed a second contact point, \code{campana}, which is a software package that subscribes to alerts and sends notifications via email, SMS, and phone calls. Users subscribe to alert groups on \code{campana}, filtering out alarms unrelated to their expertise. The four high-priority alarms described in Section~\ref{HK data} are distributed via phone calls to ensure immediate attention. \code{campana} consists of three code repositories: the server backend, the web frontend, and the core library. \par

The \code{campana} backend consists of a REST API, written using the Flask framework, that receives alert information in the form of JSON data from Grafana via the HTTP POST method. The Flask server then publishes the data to a Redis\footnote{\url{https://redis.io/}} database. We use Redis as an alert queue system, using Pub/Sub as the messaging paradigm to which the core library subscribes. Thus, Redis bridges the connection between alerts emitted by Grafana and the users receiving alerts distributed by the core \code{campana} library. \par

\setcounter{footnote}{0}

The \code{campana} frontend, shown in Figure~\ref{fig:campana}, is developed using Vue\footnote{\url{https://vuejs.org/}}. Users can enter their email addresses and phone numbers as well as subscribe to groups of alerts they wish to receive. Subscriptions are handled in Vue by toggling notification methods and pre-defined groups. This easily allow ROCs to begin their shift by turning on their notification methods. Users can turn on phone calls triggered by the most critical, time-sensitive alarms. The user address and subscription information is stored in an SQLite\footnote{\url{https://www.sqlite.org/}} database. The Flask server used for the backend is also used as the API for the web frontend; it reads the HTTP requests from the Vue interface and updates the SQLite database. \par

The core \code{campana} library consists of classes and methods to interact with the software required for alert formatting and distribution. The API is used by \code{campanad}, a \code{systemd} service that performs the above functions. As a \code{systemd} service, the alert system will function as long as the site computer running \code{campana} is operational. Each JSON-formatted alert from the Redis server contains information including what thresholds are being triggered and which groups to notify. \code{campanad} is subscribed to the Redis server and transforms the alert from JSON to text appropriate for email or SMS messaging. \code{campana} also reads the SQLite database for subscription information to send alerts to the appropriate users via the active notification methods. \code{campana} uses the Gmail SMTP server to send emails and Twilio\footnote{\url{https://www.twilio.com/en-us}} to send SMS and phone calls. Phone calls are generated from the alerts using Twilio's text-to-speech, but only contain the name of the alert that is firing to indicate which subsystem to investigate. Each notification is distributed according to the groups labeled by the alert and the methods users have activated. \par

To ensure the robustness of the alarm system, the software can be automatically restarted after unexpected shutdowns due to power outages. The \code{campana} backend and frontend run in Docker containers that are configured to start on reboot. The \code{campanad} service also emits a Slack notification if it crashes for any reason. However, network outages will cause interruptions since a connection from the site to the internet is required for distribution using the services described above. This also means that local site engineers will not receive alerts while at the site if the network is interrupted. SO recently employed a fiber connection to North America through the ALMA telescope site, and we also have a backup radio link to a low site to maintain network connectivity in the event of fiber issues. \par



\section{Deployment at Site} \label{deployment}
The SO alarm system described in Section~\ref{alarm system overview} was first tested in-lab at Yale University and is currently deployed on-site as an integral part of observatory operations. The system monitors and alerts on data from all four telescopes and the site. Each telescope uses its own computing node which hosts the \code{ocs} Agents. \cite{brian_spie, sanah_spie} The alarm system software (Grafana, InfluxDB, \code{campana}, etc) runs on a special computing node designated for site services. While the core function is the same throughout the observatory, we separate alarms into 6 overarching groups: one for the LAT, one for each of the 3 SATs, one for general site metrics, and one for data packaging/processing. Each group has its own Grafana alarm dashboard, \code{campana} notification group, and Slack notification channel. This allows easier separation of tasks for each group of SO researchers. \par

The alarm system has been operational since September 2023. Provided power and network, the system has not crashed during this time and has successfully emitted each triggering alert. The number of alert rules on Grafana has gradually increased to over 500 individual alerts at the time of writing. Many improvements have been added to the alarm system, including the addition of groups. The alert groups, which users can subscribe to using \code{campana} as described in Section~\ref{alert contact points}, help prevent notification fatigue since each telescope's team usually does not need to be aware of another's status. We also make use of Grafana's silencing feature to turn off alarms during situations such as DR cooldowns or warmups which would trigger many unnecessary alerts. Another improvement we have implemented is a feature where alerts are emitted for the observing scheduler, which is neither an \code{ocs} Agent nor part of the data processing pipeline. \cite{nextline} This allows researchers to be aware of situations quickly and not lose precious observation time. \par

\begin{figure}
    \centering
    \includegraphics[scale=0.5]{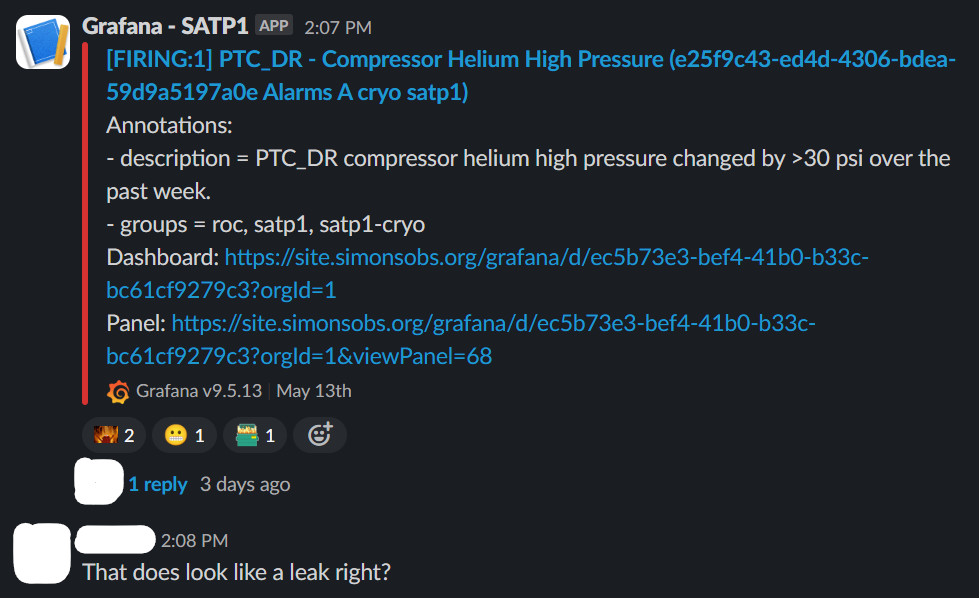}
    \caption{Slack message showing an alert that caused researchers to investigate the telescope. The alert is sent to one of the channels appointed for receiving alarms. The notification is formatted to provide context of the condition that causes the trigger. In this case, the alert was triggered when the PTC helium pressure changed by \textgreater30 psi over a week. Researchers found a leak from a high-pressure helium line for the PTC that cools the DR.}
    \label{fig:leak_alert}
\end{figure}

We have had significant situations in which alarms help save observation time and prevent irreparable damage to the telescopes. For example, schedule crash alerts have been especially useful for ROCs. Another common situation is the high disk usage alarm, as described in Table~\ref{table:alerts}. In one critical situation, there was a pressure leak in the PTC helium compressor lines. A small leak is usually difficult to discover due to the long period in which pressures may decrease. We use an alarm that triggers when PTC helium pressure changes by \textgreater30 psi over the span of a week. The Slack notification, as shown in Figure~\ref{fig:leak_alert}, notified researchers of this situation, prompting personnel to inspect the telescope to find the location of the leak. \par

We continuously add alarms as new hardware or new situations appear. At the time of writing, the deployment of data packaging and data processing alarms is in progress. We are developing our software to acquire detector metrics and create alarms to determine detector quality and analysis adequacy. We are working with the initial data processing to define the metrics from each step of the processing pipeline and continue to add visualizations to those steps. The alarm system continues to scale with the growing needs of the observatory. \par

\section{Summary} \label{summary}
We have presented an overview of the Simons Observatory alarm system, observation data processing and related data quality monitoring, and the deployment of these systems on-site. With the use of software tools such as Grafana, InfluxDB, and Slack, combined with packages written for SO such as \code{ocs} and \code{campana}, we employ a robust alarm system that allows successful observatory operations. Due to the many intricacies of a full-fledged observatory, many faults may cause loss of observing time, data corruption, and hardware damage. With the alarm system in place, we can promptly react to and prevent these issues. \par

We have presented the visualizations of the data processing pipeline, demonstrating additional capabilities for data quality monitoring. Using plots produced by \code{sotodlib} scripts running on automatic \code{prefect} schedules, we inspect the process for poor data due to internal (e.g., detector performance) and external (e.g., bad weather) factors. With metrics such as the number of detectors cut during the processing steps, we can emit alerts and catch these issues early in the pipeline. \par

As with much of the SO infrastructure, the alarm system can continue to scale up to meet the needs of efficient and satisfactory operations. This system will expand to assist with monitoring new SATs from SO:UK and SO:Japan, along with additional detectors and a renewable energy system planned for Advanced Simons Observatory (ASO). \par

\appendix
\section{ACRONYMS} \label{appendix:a}
The acronyms used in this paper are described in Table~\ref{table:acronyms}.


\section*{ACKNOWLEDGMENTS}
This work was funded by the Simons Foundation (Award \#457687, B.K.) and Yale University. 
We would like to thank the communities of the many open-source packages in use with \code{campana} and \code{sotodlib}.

\bibliography{main} 
\bibliographystyle{spiebib} 

\begin{landscape}
\begin{table}[]
\caption{Example alarms and their conditions for each alarm group. Those in red are high-severity alarms that trigger phone calls.}
\label{table:alerts}
\resizebox{\columnwidth}{!}{%
\begin{tabular}{|p{0.1\linewidth}|p{1cm}p{1cm}p{0.35\linewidth}|}
\hline
\multicolumn{1}{|c|}{\multirow{2}{*}{\textbf{Alarm Group}}} &
  \multicolumn{3}{c|}{\textbf{Examples}} \\ \cline{2-4} 
\multicolumn{1}{|c|}{} &
  \multicolumn{1}{c|}{\textbf{Metric}} &
  \multicolumn{1}{c|}{\textbf{Alarm Condition}} &
  \multicolumn{1}{c|}{\textbf{System Failure Preventions}} \\ \hline
computing &
  \multicolumn{1}{p{0.2\linewidth}|}{site computer disk usage} &
  \multicolumn{1}{l|}{\textgreater90\%} &
  A full disk causes software such as \code{ocs} Agents to crash and HK/observational data can be lost. Experts can clear disk space before this happens. \\ \hline
SMuRF &
  \multicolumn{1}{p{0.2\linewidth}|}{SMuRF FPGA current (A) and temperature (C)} &
  \multicolumn{1}{p{0.2\linewidth}|}{(\textgreater{}50A and \textgreater{}100C) or (\textgreater{}53A and \textgreater{}73C)} &
  While the SMuRFs have internal shutdown limits, these alarms allow ROCs to stop operations before hardware damage. \\ \hline
cryogenics &
  \multicolumn{1}{p{0.2\linewidth}|}{{\leavevmode\color[HTML]{FE0000} DR temperature sensor}} &
  \multicolumn{1}{p{0.2\linewidth}|}{{\leavevmode\color[HTML]{FE0000} \textgreater{}120 mK}} &
  {\leavevmode\color[HTML]{FE0000} Since the TESs need to be superconducting, poor data quality results from high temperatures. ROCs can determine DR issues before more observations continue.} \\ \hline
HWP &
  \multicolumn{1}{p{0.2\linewidth}|}{HWP spin frequency} &
  \multicolumn{1}{p{0.2\linewidth}|}{\textgreater{}3 Hz} &
  HWP frequencies higher than within spec can cause hardware damage. ROCs can spin down the HWP before this happens. \\ \hline
power &
  \multicolumn{1}{p{0.2\linewidth}|}{UPS state} &
  \multicolumn{1}{p{0.2\linewidth}|}{``on battery''} &
  During power outages, UPSs must prevent crucial hardware from losing power. Along with automatic shutdown procedures, ROCs can safely stop operations before hardware damage. \\ \hline
platform &
  \multicolumn{1}{p{0.2\linewidth}|}{ACU lockout status} &
  \multicolumn{1}{p{0.2\linewidth}|}{``platform remote control locked out''} &
  Controlled by the ACU Agent \cite{acu}, the telescope platform requires maintenance by the site crew in certain situations. For safety, ROCs cannot remotely move the telescope when the platform is locked out. \\ \hline
agents &
  \multicolumn{1}{p{0.2\linewidth}|}{data acquisition status} &
  \multicolumn{1}{p{0.2\linewidth}|}{``failed''} &
  If any Agents crash for various reasons, these alarms allow ROCs to take necessary actions to reboot them. \\ \hline
timing &
  \multicolumn{1}{p{0.2\linewidth}|}{timing device GPS sync} &
  \multicolumn{1}{p{0.2\linewidth}|}{``unsynchronized''} &
  This system is critical for timing synchronization of HK/observational data. When the central timing device becomes unsynchronized from GPS, ROCs can recognize the cause of any inconsistencies in the signal timestamps. \\ \hline
environmental &
  \multicolumn{1}{p{0.2\linewidth}|}{{\leavevmode\color[HTML]{FE0000} site wind speed}} &
  \multicolumn{1}{p{0.2\linewidth}|}{{\leavevmode\color[HTML]{FE0000} gusts \textgreater{}70 km/hr}} &
  {\leavevmode\color[HTML]{FE0000} While the site crew is aware of wind speed, ROCs must also be informed since the telescope platform is rated to move within certain conditions. This allows ROCs to interrupt observations before hardware damage.} \\ \hline
\end{tabular}%
}
\end{table}
\end{landscape}

\begin{table}[h!]
\centering
\caption{Acronyms.}
\label{table:acronyms}
\begin{tabular}{ |l|l| }
 \hline
 \textbf{Acronym}&\textbf{Definition}\\
 \hline
 ACU&Antenna Control Unit\\
 API&Application Programming Interface\\
 ASO&Advanced Simons Observatory\\
 CMB&Cosmic microwave background\\
 CPU&Central processing unit\\
 DR&Dilution refrigerator\\
 FPGA&Field Programmable Gate Array\\
 HK&Housekeeping\\
 HTTP&Hypertext Transfer Protocol\\
 HWP&Half-wave plate\\
 IRIG&Inter-range instrumentation group timecodes\\
 JSON&JavaScript Object Notation\\
 LAT&Large-aperture telescope\\
 \code{ocs}&Observatory Control System\\
 PTC&Pulse Tube Cryocoolers\\
 PTP&Precision Time Protocol\\
 PWV&Precipitable water vapor\\
 REST&Representational State Transfer\\
 ROC&Remote Observing Coordinator\\
 SAT&Small-aperture telescope\\
 SMS&Short Message Service\\
 SMTP&Simple Mail Transfer Protocol\\
 SMuRF&SLAC Microresonator Radio Frequency\\
 SO&Simons Observatory\\
 TES&Transition edge sensor\\
 $\mu$MUX&Microwave multiplexing\\
 UPS&Uninterruptible Power Supply\\
 \hline
\end{tabular}
\end{table}

\end{document}